\documentclass[preprint,aps]{revtex4}
\usepackage{mathrsfs}

\usepackage{graphicx}
\usepackage{multirow}
\usepackage{makecell}
\usepackage{booktabs}

\usepackage{dcolumn}
\usepackage{amsmath, bm}
\usepackage{xcolor}
\usepackage{wasysym}
\usepackage{amssymb}

\newcommand{\ef}{$E_F$}

\newcommand{\kz}{$k_z$}
\newcommand{\kx}{$k_x$}

\newcommand{\dxxyy}{$d_{x^2-y^2}$}
\newcommand{\dz}{$d_{z^2}$}

\newcommand{\dxy}{$d_{xy}$}
\newcommand{\g}{$\Gamma$}
\newcommand{\crr}{CeRu$_2$}
\newcommand{\beq}{\begin{eqnarray}}
\newcommand{\eeq}{\end{eqnarray}}
\def \bs{\textbf}
\newcommand{\yi}[1]{\textcolor{blue}{#1}}

\begin{document}



\title{Observation of Flat Bands and Dirac Cones in a Pyrochlore Lattice Superconductor}

\author{
\parbox{\linewidth}{\centering
Jianwei Huang$^{1}$, Chandan Setty$^{1}$, Liangzi Deng$^{2}$, Jing-Yang You$^{3}$, Hongxiong Liu$^{4}$, Sen Shao$^{5}$, Ji Seop Oh$^{6,1}$, Yucheng Guo$^{1}$, Yichen Zhang$^{1}$, Ziqin Yue$^{1}$, Jia-Xin Yin$^{7}$, Makoto Hashimoto$^{8}$, Donghui Lu$^{8}$, Sergey Gorovikov$^{9}$, Pengcheng Dai$^{1}$, Jonathan D. Denlinger$^{10}$, M. Zahid Hasan$^{7,11}$, Yuan-Ping Feng$^{3,12}$, Robert J. Birgeneau$^{6,13,14}$, Youguo Shi$^{4}$, Ching-Wu Chu$^{2,13}$, Guoqing Chang$^{5}$, Qimiao Si$^{1}$, Ming Yi$^{1,*}$\\}
}

\affiliation{
$^{1}$Department of Physics and Astronomy, Rice Center for Quantum Materials, Rice University, Houston, Texas 77005, USA\\
$^{2}$Department of Physics and Texas Center for Superconductivity, University of Houston, Houston, Texas 77204, USA\\
$^{3}$Department of Physics, National University of Singapore, 2 Science Drive 3, Singapore 117551, Singapore\\
$^{4}$Beijing National Laboratory for Condensed Matter Physics, Institute of Physics, Chinese Academy of Sciences, Beijing 100190, China\\
$^{5}$Division of Physics and Applied Physics, School of Physical and Mathematical Sciences, Nanyang Technological University, Singapore 637371, Singapore\\
$^{6}$Department of Physics, University of California Berkeley, Berkeley, California 94720, USA\\
$^{7}$Laboratory for Topological Quantum Matter and Advanced Spectroscopy (B7), Department of Physics, Princeton University, Princeton, New Jersey 08544, USA\\
$^{8}$Stanford Synchrotron Radiation Lightsource, SLAC National Accelerator Laboratory, Menlo Park, California 94025, USA\\
$^{9}$Canadian Light Source, 44 Innovation Boulevard, Saskatoon SK S7N 2V3, Canada\\
$^{10}$Advanced Light Source, Lawrence Berkeley National Laboratory, Berkeley, California 94720, USA\\
$^{11}$Princeton Institute for Science and Technology of Materials, Princeton University, Princeton, New Jersey 08544, USA\\
$^{12}$Centre for Advanced 2D Materials, National University of Singapore, 6 Science Drive 2, Singapore 117546, Singapore\\
$^{13}$Materials Sciences Division, Lawrence Berkeley National Laboratory, Berkeley, California 94720, USA\\
$^{14}$Department of Materials Science and Engineering, University of California, Berkeley, California 94720, USA\\
$^{*}$To whom correspondence should be addressed: mingyi@rice.edu
}

\date{\today}

\begin{abstract}
Emergent phases often appear when the electronic kinetic energy is comparable to the Coulomb interactions. One approach to seek material systems as hosts of such emergent phases is to realize localization of electronic wavefunctions due to the geometric frustration inherent in the crystal structure, resulting in flat electronic bands. Recently, such efforts have found a wide range of exotic phases in the two-dimensional kagome lattice, including magnetic order, time-reversal symmetry breaking charge order, nematicity, and superconductivity. However, the interlayer coupling of the kagome layers disrupts the destructive interference needed to completely quench the kinetic energy. Here we demonstrate that an interwoven kagome network---a pyrochlore lattice---can host a three dimensional (3D) localization of electron wavefunctions. Meanwhile, the nonsymmorphic symmetry of the pyrochlore lattice guarantees all band crossings at the Brillouin zone X point to be 3D gapless Dirac points, which was predicted theoretically but never yet observed experimentally. Through a combination of angle-resolved photoemission spectroscopy, fundamental lattice model and density functional theory calculations, we investigate the novel electronic structure of a Laves phase superconductor with a pyrochlore sublattice, \crr. We observe flat bands originating from both the Ce 4$f$ orbitals as well as from the 3D destructive interference of the Ru 4$d$ orbitals. 
We further observe the nonsymmorphic symmetry-protected 3D gapless Dirac cones at the X point. Our work establishes the pyrochlore structure as a promising lattice platform to realize and tune novel emergent phases intertwining topology and many-body interactions.

\end{abstract}

\maketitle

\newpage

\section{Introduction}
A major recent effort in the field of quantum materials is to engineer systems where topological fermion physics is realized in the regime of strong electron correlations. This effort has largely been explored in creating mini flat bands in twisted bilayer graphene~\cite{Bistritzer2011,Cao2018} and transition metal dichalcogenides~\cite{Regan2020,Tang2020}, topologically protected band crossings with correlation-driven narrow bandwidth in Weyl-Kondo systems~\cite{Lai2018,Dzsaber2021}, and the coexistence of Dirac crossings and flat bands in geometrically frustrated systems~\cite{Henley2010,Tang2011,Sun2011,Neupert2011,Liu2014b,Derzhko2015,Kang2020b}. Flat bands in momentum space correspond to localized states in real space, where the quenched kinetic energy of electrons is negligible in comparison to the Coulomb interactions. In such a regime, many-body effects can lead to emergent phases such as magnetism, density waves, nematicity, and superconductivity~\cite{Tasaki1992,Sharpe2019,Liu2018b,Cao2021,Nie2022,Li2022,Miyahara2007,Cao2018}. Arguably the system with the most abundant material platforms realized amongst these efforts so far are the kagome lattice compounds~\cite{Ye2018,Ortiz2019a,Kang2020c,Ortiz2021a,Yin2022,Setty2022,Teng2022}. Consisting of corner-sharing triangles, the two-dimensional (2D) kagome lattice has been well-known as one of the most promising candidates for hosting quantum spin liquid states arising from the extensively degenerate magnetic ground states due to the lattice geometry-induced magnetic frustration~\cite{Balents2010}. Analogously, fermionic models on kagome lattices have also predicted the existence of electronic flat bands arising from a geometry-induced destructive interference of the electronic wavefunction~\cite{Henley2010,Tang2011,Sun2011,Neupert2011}. In addition, the point group of kagome is the same as that of graphene, therefore also guaranteeing the presence of Dirac fermions in the electronic structure~\cite{CastroNeto2009}. However, theoretical models for kagome lattices are largely 2D. In real material systems recently realized, kagome lattices are stacked in the bulk crystal structure where interlayer coupling is finite, giving rise to non-negligible dispersion along the \kz~direction, where the bandwidth of the flat bands could be on the order of hundreds of meV~\cite{Liu2020,Huang2020}.

Pyrochlore lattice, made of corner-sharing tetrahedrons, can be considered as a three dimensional (3D) analog of the kagome lattice, where geometric frustration is naturally realized in 3D (Fig. 1b). Similarly, pyrochlore crystals have been proposed and widely studied as candidates to host quantum spin liquid states from the 3D magnetic frustration in the past~\cite{Anderson1956,Hermele2004,Savary2012,Sibille2015,Gao2019,Gaudet2019}. In the fermionic models, the same geometric frustration has been theoretically predicted to lead to bands that are flat along every direction in the 3D momentum space due to quantum destructive interference~\cite{Essafi2017,Hase2018}. Meanwhile, previous theory studies have also predicted four-fold degenerated Dirac point at the high symmetry point of the Brillouin zone in the pyrochlore lattice, which, with further reduced symmetry, could evolve into Weyl points or gapped Dirac point~\cite{Fu2007,Young2012,Wan2011a}. The Dirac point is protected by the crystalline nonsymmorphic symmetry and fundamentally different from the previously observed Dirac cones induced by band inversions~\cite{Wang2012c,Yang2014b,Liu2014c}. However, there exists no direct experimental observation of neither the destructive interference-induced flat band nor the nonsymmorphic symmetry-protected Dirac point in the pyrochlore lattice compound.

In this study, through a combination of angle-resolved photoemission spectroscopy (ARPES), theoretical modeling, and density functional theory (DFT) calculations, we report the electronic structure of a Laves phase superconductor, CeRu$_2$, that possesses a Ru pyrochlore sublattice. In addition to the flat bands originating from the Ce 4$f$ states, we observe flat bands of Ru 4$d$ orbitals, attributed to the 3D destructive interference inherent in the pyrochlore structure. The flat bands are actually partially flat in the momentum space due to strong hybridization with the Ce 4$f$ orbitals, but still contribute to a large peak in the electronic density of states (DOS). In addition, we also observe 3D gapless Dirac crossings protected by the nonsymmorphic symmetry of the pyrochlore structure. Our results establish the metallic pyrochlore systems to be a new platform for hosting the rich phenomenology of correlated topology.

\section{Results}
While studies of insulating pyrochlores are abundant in the context of quantum spin liquids, limited ARPES studies of metallic pyrochlores are reported due to the challenge in cleaving the 3D crystals~\cite{Kondo2015,Wang2015,Nakayama2016}. The recent improvement of beamspot focusing has enabled exploration of a wider range of 3D materials. Here we identify \crr~as an ideal candidate to explore the pyrochlore lattice as a platform for realizing correlated topology. The crystal structure of \crr~is that of the Laves phase in space group 227 (Fd$\overline3$m) (Fig. 1a)~\cite{Palenzona1991}. The Ce atoms form a diamond sublattice (Wyckoff position 8 (a)) while the Ru atoms form the pyrochlore sublattice (Wyckoff position 16 (d)) as shown in Fig. 1b. 

\crr~was first discovered in the search for magnetic superconductors over six decades ago~\cite{Matthias1958}. It has a superconducting transition temperature around 5 K (Fig. 1c and d). Previous angle-integrated photoemission spectroscopy measurements have revealed an $s$-wave superconducting gap~\cite{Kiss2005}.
Recent high pressure work shows superconductivity to persist up to at least 168 GPa, with an appearance of a secondary superconducting transition onsetting at 30 GPa~\cite{Deng2022}. The superconducting state is unusual, marked by a secondary magnetic hysteresis loop, which originally led to interpretation of reentrant superconductivity but later understood to be due to a Fulde-Ferrell-Larkin-Ovchinnikov state~\cite{Deng2022}. Evidence for weak magnetism has also been suggested by Muon spin rotation ($\mu$SR) experiments in zero magnetic field~\cite{Huxley1996, Mielke2022}.  There is, however, no corresponding magnetic transition in macroscopic magnetic susceptibility measurements (see Supplementary Note 1 and Supplementary Figure 1). Polarized neutron scattering has found field-induced paramagnetic moment on the order of $10^{-4}\mu_BT^{-1}$ interpreted to be equally distributed on both the Ce and Ru sites~\cite{Huxley1997}. Further clarification, however, is needed on the origin of magnetism in \crr. Interestingly, previous photoemission measurements of \crr~have identified two peak structures near the Fermi level in the integrated density of states, interpreted as to originate from the Ce 4$f$-electron states~\cite{Yang1996,Kang1999,Sekiyama2000}. However, no momentum-resolved measurement of the electronic structure has been reported. 

\subsection{Tight-binding model}
Before we present our ARPES experimental observation, to understand the salient features of the electronic structure of a pyrochlore lattice, we present calculations from a tight-binding model (see Methods for details), the results of which are consistent with previous studies~\cite{Bergman2008,Guo2009}. First in the absence of spin-orbit coupling (SOC), the nearest-neighbor hopping yields four bands that include a pair of exactly flat degenerate bands, a Dirac nodal line between the X and W points in the Brillouin zone (BZ), and a triple point fermionic crossing at the $\Gamma$ point (see Fig. 1f). Here we note that the pyrochlore lattice is the line-graph lattice of the diamond lattice. Hence the flat bands here are characteristics of the line-graph nature of the pyrochlore lattice~\cite{Kollar2020}. To understand the origin of the flat bands more intuitively, we look at two cut planes of the pyrochlore lattice, each of which is a 2D kagome plane formed by the Ru atoms (Fig. 1h). These kagome planes can be thought of as the origin of the degenerate pair of flat bands. To see this clearly, we write the normalized eigenstates of the two degenerate flat bands as 
\begin{eqnarray} \label{v1}
\textbf v_1(k_x, k_y, k_z) = 
\frac{1}{\sqrt{N_1}}\left( \begin{array}{cccc}
 s_{k_x - k_z},   
 s_{k_y + k_z}, 
0, 
- s_{k_x + k_y} 
\end{array}
\right)
\end{eqnarray}
\begin{eqnarray} \label{v2}
\textbf v_2(k_x, k_y, k_z) = 
\frac{1}{\sqrt{N_2}}\left( \begin{array}{cccc}
 s_{k_y - k_z},   
 s_{k_x + k_z},
-s_{k_x + k_y},  
0 
\end{array}
\right)
\end{eqnarray}
where the normalization factors are  $N_1 =  s_{k_x - k_z}^2 +  s_{k_y + k_z}^2 + s_{k_x + k_y}^2$ and $N_2 = s_{k_y - k_z}^2 + s_{k_x + k_z}^2 + s_{k_x + k_y}^2$ with the definition $s_{k_i \pm k_j} \equiv \sin(k_i \pm k_j) $. The Wannier orbitals associated with the two degenerate eigenstates (Eqs. 1, 2) are shown in Fig. 1h. In the chosen basis, it is clear from the structure of Eqs. 1, 2 how the two Wannier orbitals decompose into decoupled kagome planes -- the eigenvector $\textbf v_1 (\textbf v_2)$ has zero Wannier amplitude in the third (fourth) sublattice component. This effectively `disconnects' the pyrochlore lattice into 2D kagome planes along two distinct directions as the out-of-plane Wannier amplitudes at all sites connecting the parallel kagome planes are \textit{identically zero}. Within each kagome plane, the flat band arises from a destructive interference of the remaining three sinusoidal Wannier amplitudes in Eqs. 1, 2, with phase factors alternating in sign on the 2D kagome hexagon (see Fig. 1h). Thus the mechanism for flat band formation in the 3D pyrochlore lattice can be mapped to that of (two copies of) the 2D kagome problem. \par
With the incorporation of SOC (Fig. 1g), the triply-degenerate point at $\Gamma$ will be lowered to form a singly-degenerate point and a doubly-degenerate quadratic band touching node. Whether or not the latter involves the dispersive bands depends on the sign of the SOC, $\lambda$. For a negative value of $\lambda$, the doubly-degenerate subspace is formed entirely by the two flat bands, whereas for a positive value of $\lambda$, it involves one of each of the dispersive and flat bands.
More interestingly, The double degeneracy along X -- W will be lifted with only the degeneracy remaining at the X point to form a 3D Dirac point. This was the first proposed 3D Dirac point protected by nonsymmorphic symmetry but has never yet been observed experimentally~\cite{Fu2007,Young2012}. In the following sections, we will present the unambiguous observation of the two signatures of the pyrochlore lattice that establishes it as a topological geometrically frustrated metal -- the destructive interference-induced flat bands and the nonsymmorphic symmetry-protected Dirac cones.

\subsection{Electronic structure and Ce 4$f$ flat bands}
We first present the overall measured band structure of \crr. In Fig. 2a and b we present the out-of-plane and in-plane Fermi surface mappings of the \crr~single crystal with respect to the (111) cleaved surface, respectively. The corresponding BZs are also overlaid for reference. The in-plane Fermi surface mapping (Fig. 2b) shows a mirror symmetry about the vertical axis. The three-fold rotational symmetry of the (111) surface can be clearly seen in the three point-like spots within the BZ as well as the larger pockets outside the first BZ. The out-of-plane mapping is carried out via a photon-energy dependence measurement of the cut shown in the inset of Fig. 2a. The periodicity along \kz~can be seen in the out-of-plane mapping at higher zones centered at \kx~$\sim$ 3 \AA$^{-1}$ (Fig. 2a). To gain more insights into the band dispersions, we show the energy-momentum spectrum measured with 47 eV photons (Fig. 2c), which corresponds to the L -- \g~-- L direction as indicated in Fig. 2a. The band dispersions by DFT calculations along this high symmetry direction are plotted in Fig. 2d for comparison. We note that we do not see any evidence for magnetic band splitting within our experimental resolutions at temperatures below 40 K (see Supplementary Note 3 and Supplementary Figure 5) and therefore we compare our data with the results of nonmagnetic calculations exclusively (see Supplementary Figure 3 and 4 for comparison of nonmagnetic and magnetic calculations). Several dispersive hole bands are observed near \ef~centered at the \g~point, which are reproduced by the DFT calculations. The less dispersive bands near -2.5 eV in the data are also nicely captured by the calculations with a renormalization factor of 1.2. In general, the ARPES data and the DFT calculations exhibit a good consistency.

In the Ce-based materials, a common characteristic of the electronic structure is the presence of two flat bands. One is the Kondo resonance state (4$f^1_{5/2}$) located at the Fermi level, and the other is its spin-orbit sideband (4$f^1_{7/2}$) below \ef. DFT calculations have been shown to be unable to accurately reproduce these two flat bands~\cite{Fujimori2016,Denlinger2001}. In Fig. 2e, we plot the spectral image measured with 120 eV photons (Ce 4$d$$\rightarrow$4$f$ resonance) under linear horizontal polarization. The corresponding momentum path is illustrated in Fig. 2a. Here, two flat bands are clearly discernible, with one near the Fermi level and one at around -0.25 eV. These two flat bands are also evident in the peaks in the energy distribution curve (EDC) stacks (Fig. 2f). They correspond to the Ce 4$f^1_{5/2}$ and 4$f^1_{7/2}$ states (Detailed photon energy-dependent measurements across the resonance energies of the Ce 4$f$ states are shown in Supplementary Figure 6 and 7), which were also observed as two peaks in the previous angle-integrated photoemission reports~\cite{Yang1996,Sekiyama2000}. The spectra of Ce 4$f$ flat bands observed here have a predominantly surface origin due to the short photoelectron mean free path associated with the utilized photon energy (120 eV)~\cite{Allen1983,Yang1996,Sekiyama2000}. The presence of these Ce 4$f$ states poses a challenge for observing flat bands induced by destructive interference of the Ru sublattice. In the following, to unambiguously reveal the flat bands originating from the pyrochlore lattice, we first identify the corresponding destructive-interference-induced flat bands from the calculations, and then demonstrate their existence via ARPES measurements using a photon energy where the Ce 4$f$ spectral intensity is suppressed (see also Supplementary Note 4 and Supplementary Figure 7)~\cite{Fujimori2016,Denlinger2001}.

\subsection{Observation of flat bands induced by destructive interference}
To identify the flat bands in \crr~originating from the pyrochlore structure, we performed a series of systematic band structure calculations on Ru$_2$, LaRu$_2$, and finally \crr~(Fig. 3a-d). First, we examine the band structure of the artificial bare Ru$_2$ structure, which is obtained by removing all Ce atoms from \crr, resulting in a Ru$_2$-only pyrochlore lattice (inset in Fig. 3a). The pyrochlore flat bands are most distinctly visible for the Ru$_2$-only pyrochlore lattice (shaded region in Fig. 3a), positioned just above \ef~with Ru 4$d$ orbital character. The 3D pyrochlore flat bands are observed to slightly deviate from the simple tight-binding model of an extremely flat band extending throughout the entire BZ due to strong hybridization with a dispersive band along \g--X. To see how these flat bands evolve ultimately to those in \crr, we next calculate the band structure of LaRu$_2$, which is isostructural to \crr~but with no $f$-electrons near \ef(Fig. 3b). This is followed by \crr~without SOC (Fig. 3c), and then SOC is finally added (Fig. 3d).
We track the evolution of the Ru $d$-originated flat bands by showing the side by side comparison of the band calculations projected on the Ru-$d$ orbitals along this series of calculations. Two main effects are evident. First, a downward shift of the Ru-$d$ orbital bands can be seen, which can be attributed to the La/Ce sites acting as electron donors to Ru sites. Second, the strong hybridization of the Ru electronic states and the La/Ce electronic states further widens the bandwidth of the flat bands and results in a more restricted k-space range of the flat bands. Consequently, with the inclusion of SOC, we identify the flat bands in the band calculations of \crr~to be mainly located around -0.6 eV with a much broadened band dispersion (shaded region in Fig. 3d). Despite the finite bandwidth, these bands can still exhibit significant flatness along specific directions in momentum space. We note that part of the flat band is segmented into a flat portion within a small k-region around the $\Gamma$ point (blue arrow in Fig. 3d). A complementary view of the energy shift and broadening effect of the flat band dispersion is provided by a comparison plot of the Ru-$d$ partial DOS in Fig. 3e. The presence of a similar Ce 4$f$ DOS peak, also at -0.6 eV, provides evidence for the strong hybridization of the Ce 4$f$ states with the Ru 4$d$ flat band states.

Having theoretically identified the pyrochlore flat bands in \crr, we now present the ARPES data. To minimize spectral intensity from Ce 4$f$ orbitals, we utilize photons with energy lower than the Ce 4$d$$\rightarrow$4$f$ resonance of 120 eV. We present a spectral image along L--$\Gamma$--L in the near-\ef~region measured using a linear vertical polarization at 47 eV (Fig. 3g). Several hole-like dispersive bands are clearly resolved, consistent with predictions from the DFT calculations (Fig. 3d and Fig. 2d). In addition, we observe almost nondispersive features around -0.25 eV and -0.5 eV, which are also revealed distinctly by the corresponding EDC stacks (Fig. 3h). The flat band at -0.25 eV was identified as the Ce 4$f^1_{7/2}$ states, with the intensity strongly suppressed. The flat feature at around -0.5 eV has no other correspondence but aligns with the pyrochlore flat bands predicted by the DFT calculations at around -0.6 eV (Fig. 3d). The observed band appears to be flatter than in the calculation. This is likely due to the broad linewidth of the observed spectra. We note that the DOS peak at around -0.5 eV in \crr~has been previously reported in angle-integrated photoemission measurements, correlated to the -0.5 eV peaks in experimental off-resonance (Ru-$d$) spectra \cite{Kang1999,Sekiyama2002}, and in Ce 3$d$$\rightarrow$4$f$ resonance spectra \cite{Sekiyama2002}. Our results unequivocally establish that the pyrochlore flat bands are the underlying origin of the theoretical and experimental DOS peaks in \crr. Furthermore, the flat band segment at -0.2 eV in the calculation around the BZ center (Fig. 3d) is clearly observed with a helium lamp using 21.2 eV photons (Fig. 3i), where the two Ce 4$f$ states in the low energy are completely absent due to a low Ce 4$f$ cross section~\cite{Garnier1997}. This flat band segment is observed to be at -0.11 eV and corresponds to a very sharp peak in the integrated EDC (Fig. 3j). It can be brought even closer to the Fermi level with hole doping in \crr. Taking all the measurements and band calculations together, we have successfully identified the flat bands induced by 3D destructive interference of the Ru pyrochlore sublattice in \crr. 

\subsection{Observation of 3D Dirac cones}
Next, we explore the nontrivial band topology of \crr. The space group of \crr~is Fd$\overline3$m, from which the first Dirac semimetal and Weyl semimetal were theoretically proposed~\cite{Fu2007,Young2012,Wan2011a}. The crystal structures that belong to this space group include diamond lattice, pyrochlore lattice, $\beta$-cristobalite and spinel oxides. The same reasoning applies to the protection of the Dirac point at X for these compounds, in which the nonsymmorphic symmetries play an essential role. Specifically, for the pyrochlore lattice, the combination of two nonsymmorphic operations $\{I|t_{0}\}$, $\{C_{4x}|t_{0}\}$ and one symmorphic operation $\{C_{2z}|0\}$ protects the Dirac point at X. Here $I$ is the inversion operation, $C_{4x}$ is the four-fold rotation about the $x$-axis, $C_{2z}$ is the two-fold rotation about the $z$-axis, and the center of all the symmetry operations is the center of any tetrahedron. $t_{0}$ is a non-Bravais translation along the diagonal direction (Fig. 4a)~\cite{Bzdusek2015}. 
The pyrochlore lattice is in analogy to the well-studied diamond lattice. The diamond lattice can be simply viewed as formed by the centers of the tetrahedra of the pyrochlore lattice (Fig. 4b). One particular nonsymmorphic symmetry of the diamond lattice (Fig. 4c)~\cite{Grefe-Si2020} is the screw axis $\{C_{4z}|t_{c/4}\}$, where $C_{4z}$ is the four-fold rotation around axis $z$ and $t_{c/4}$ is a quarter translation along $c$. All the lattices that belong to space group Fd$\overline3$m carry four dimensional irreducible representations at the X point of the BZ~\cite{Programs1966,Young2012}. As a result, the four bands are degenerate at the X point and disperse linearly along all directions around X, forming a 3D Dirac cone. \yi{It is important to point out that, the nonsymmorphic symmetry here dictates that ALL band crossings at the X point, regardless of their energy or orbital character, are gapless 3D Dirac cones. This is clearly seen in the series of calculations shown in Fig. 3. Regardless of the orbital characters and how they hybridize, all band crossings at the X point are 3D Dirac points (see also Supplementary Note 5 and Supplementary Figure 8 for DFT calculation projected unto different orbitals). This means that it is very easy to obtain gapless Dirac points near the Fermi level in a material with a pyrochlore sublattice. We also note that comparing Fig. 3c and 3d, the effect of the SOC is to split the Dirac nodal lines along X-W, as captured by the tight-binding model.} When time-reversal symmetry is broken, the Dirac point will split into two Weyl points~\cite{Wan2011a}. This Dirac point at the BZ high symmetry point protected by nonsymmorphic symmetry (Fig. 4b) is fundamentally different from the previously observed Dirac cones induced by band inversions~\cite{Wang2012c,Yang2014b,Liu2014c} and has never yet been observed. 

Here we present the evidence for the 3D Dirac cone structure through ARPES in \crr. In Fig. 4d we plot the band image measured with 68 eV photons which corresponds to the L -- X momentum path as indicated in Fig. 4e. We can clearly observe two bands that cross at the X point to form a Dirac point at -1.7 eV (DP1). A second Dirac point with weaker intensity can also be seen around -2.5 eV (DP2). The two Dirac points can be better visualized through the second energy derivative spectral image shown in Fig. 4h, as well as by tracing the peaks in the EDC and momentum distribution curve (MDC) stacks (Fig. 4j, k). From the DFT calculations shown in Fig. 4i, we can also confirm that ALL band crossings at the X point, regardless of their energies, are indeed four-fold degenerate. Moreover, we can identify two Dirac crossings that correspond to our measured observable Dirac crossings at similar energies after considering band renormalization (Fig. 4i). The existence of the 3D Dirac cones is further verified by our near-normal emission data (Fig. 4f and g). The observation of a Dirac point around -1.7 eV is consistent with the energy position of DP1 (more details in Supplementary Figure 8). We also note that from the spectral image taken in the first BZ, we observe a clear intensity asymmetry of the two bands giving rise to the crossing (Fig. 4d, f). This is an inherent property of the nonsymmorphic symmetry protection \cite{Wu2022}, where the two bands across the BZ boundary exhibit opposite parity due to the nonsymmorphic operation. Here we note that while we present two resolvable Dirac crossings, all band crossings at the X point must be Dirac crossings, including those near \ef. Hence it would be interesting to further explore the tuning of these Dirac cones to \ef~via doping, as well as topological phase transitions with additional reduction of crystal symmetries.

\section{Discussion}
\crr~belongs to a large class of materials with the Laves phase that have the chemical composition $AB_2$, where the $B$ sublattice forms the pyrochlore structure. Superconductivity is commonly found in these compounds including SrIr$_2$, Hf$_{1-x}$Zr$_x$V$_2$ and NbBe$_2$ etc~\cite{Lawson1978,Horie2020,Rahaman2016}. The 3D destructive interference originating from the B site pyrochlore sublattice promotes flat bands in the electronic structure of those Laves phase materials. In particular, our ARPES experiments have unambiguously established that the electronic structure of \crr~consists of the Ru 4$d$ orbital flat bands close to \ef~around the BZ center. We noted that superconductivity with a similar $T_c$ was also found in the isostructural compounds SrRh$_2$ and SrIr$_2$ without $f$-electrons~\cite{Matthias1957}. Very interestingly, it has been reported that the superconducting transition temperatures of \crr~compounds by chemical substitutions on the Ce sites with La, Nd and Lu all exhibit an initial increase of $T_c$ before a monotonic decrease, while substitution by Th shows a monotonic decrease~\cite{Roy1990}. This dichotomy of behaviors are consistent with a doping-induced shift of the flat bands relative to the Fermi level. As Ce usually possesses fluctuating valence between 3+ and 4+, La, Nd and Lu all as 3+ would introduce hole doping that initially shifts the flat band closer to \ef~(also shown in the band calculations in Fig. 3b and c) before shifting further away above \ef. On the other hand, Th is 4+ and hence would electron dope \crr, which brings the flat band away from \ef. This doping-induced tuning of the flat band seems to directly correlate with the behavior of $T_c$ in this series of materials. Our studies suggest that superconductivity of the Laves phase superconductors represented by \crr~may be closely related to the topological flat bands intrinsic to the pyrochlore lattice.

Both electronic features that we have observed here in \crr, namely the destructive interference-induced flat bands and the 3D Dirac cones are guaranteed signatures of the pyrochlore lattice structure, and should appear in a wide range of pyrochlore metals including the pyrochlore iridates~\cite{Kondo2015,Wang2015}. \crr~as we demonstrated here, has its flat bands in very close proximity to \ef, and therefore less challenging to observe. The 3D nature of the pyrochlore lattice makes it fundamentally distinct from the kagome lattice. As alluded to earlier, nonsymmorphic symmetries play a crucial role in determining its electronic structure. The Dirac point is hence robust to spin mixing from SOC effects. In contrast, the Dirac cones in a 2D kagome lattice are gapped out by SOC. Additionally, the direction of the SOC is irrelevant to the band structure in  kagome systems, whereas, in the pyrochlore lattice, Fig. 1f shows how the 3D nature of the SOC makes the bands sensitive to the direction of the SOC. Moreover, the 3D nature of the pyrochlore lattice allows for strong topological insulating phases to be realized~\cite{Guo2009}. Such phases are not possible in a 2D kagome network. Furthermore, the pyrochlore electronic structure has features absent in a kagome lattice. These include triplet point crossings arising from the doubly-degenerate flat bands without SOC and the appearance of Dirac nodal lines in the limit of weak SOC. Finally, the three dimensionality of the pyrochlore lattice cuts-off any divergence in the density of states from the saddle points. This is unlike the case of kagome lattice where Van Hove singularities are expected to play a greater role in driving ordered phases~\cite{Wang2013}. Hence, the possibility of ordered phases resulting from weak coupling nesting effects is less likely; rather, the flat bands near the Fermi level serve as an ideal platform to realize correlated superconducting orders obtained from strong coupling approaches. 
Our results experimentally establish the potential for pyrochlores as a novel lattice platform with a wide range of material systems to realize and tune exotic correlated and topological phases that are to be discovered. 

\section{Methods}
\subsection{Sample growth and characterization}
The single crystal of \crr~was grown using the Czochralski pulling method in a tetra-arc furnace. Firstly, a polycrystalline ingot of mass 10 g was prepared in an arc furnace as a precursor. The ingot was then placed on the water-cooled copper hearth of the tetra-arc furnace under Ar atmosphere with a Ti ingot adsorbing oxygen. An electric current of 16 A was applied into the ingot to melt it and the hearth rotated at a speed of 0.3 revolutions per minute. A columnar single crystal was successfully grown after several hours. Electrical transport and magnetization measurements were performed using a Quantum Design Physical Property Measurement System (PPMS) and Magnetic Property Measurement System (MPMS) under magnetic fields up to 7 T and at temperatures down to 1.8 K (Supplementary Fig. S1).

\subsection{Angle-resolved photoemission spectroscopy (ARPES)}
ARPES experiments were performed at beamline 5-2 of the Stanford Synchrotron Radiation Lightsource with a DA30 electron analyzer. The angular resolution was set to 0.1$^\circ$. The overall energy resolution was set to 15 meV. The energy resolution for the temperature dependent measurements was set to 11 meV. The samples were cleaved \textit{in-situ} along the (111) crystalline plane below 15 K and the measurements were conducted in ultra-high vacuum with a base pressure of 3 $\times$ 10$^{-11}$ Torr. The soft X-ray (500 $\sim$ 880 eV) ARPES measurements were performed at the QMSC beamline of the Canadian Light Source with an R4000 electron analyzer with an energy resolution of around 100 meV. All the measurements were conducted at around 15 K unless otherwise noted. The ARPES measurements using helium lamp were performed at a lab-based system with a DA30 electron analyzer and the energy resolution was set to around 12 meV.

\subsection{Density functional theory calculations}
The first-principles calculations were performed by the Vienna Ab initio Simulation Package (VASP)~\cite{Kresse1994,Kresse1996}, within the projector augmented-wave method~\cite{Blochl1994}. The generalized gradient approximation (GGA) with the Perdew-Burke-Ernzerhof (PBE)~\cite{Perdew1996} realization was adopted for the exchange-correlation functional. The 4$f^1$5$d^1$6$s^2$ and 4$p^6$4$d^7$5$s^1$ were treated as valence electrons for Ce and Ru atoms, respectively. A plane-wave cutoff energy of 500 eV and a MonkhorstPack k-point mesh~\cite{Monkhorst1976a} with a size of 15×15×15 were used in the geometry optimization and the simulations of electronic properties. To account for the correlation effects for transition-metal elements, the DFT + U method with U = 1 eV for the 4$d$ orbital of Ru atom was used for calculating the band structures, which shows magnetism consistent with the previous experiments~\cite{Huxley1997}. Similar GGA-PBE WIEN2k \cite{Blaha2020} calculations were performed for the comparison between Ru$_2$, LaRu$_2$ and CeRu$_2$ band structures. 

\subsection{Tight-binding model}
\textit{Symmetries of the Hamiltonian:} The pyrochlore lattice belongs to space group 227 with Wyckoff positions at 16 (c) or 16 (d).  We choose primitive vectors of the FCC lattice as $\bs a_1 = (2, 2, 0)$,  $\bs a_2 = (2, 0, 2)$,  $\bs a_3 = (0, 2, 2)$.  The unit cell contains four atoms that are translated along the diagonals of the faces of a cube to form corner sharing tetrahedra. The coordinates of the four atoms are  $\bs A_1 = (0, 0, 0)$, $\bs A_2 = (1, 1, 0)$, $\bs A_3 = (1, 0, 1)$, $\bs A_4 = (0, 1, 1)$. The pyrochlore lattice contains corner sharing tetrahedra, and the point group of the lattice contains (i) a two fold rotation about an axis that passes through the midpoint of each edge of the tetrahedron and its center, (ii) a three fold rotation about an axis that passes through the vertex of the tetrahedron and its center, (iii) a four fold improper rotation about an axis that passes through the midpoint of each edge of the tetrahedron and its center, and (iv) a mirror reflection about a plane contain one edge of the tetrahedron and its center. \par
\textit{Total Hamiltonian}: The total Hamiltonian in the absence of interactions takes the form 
\begin{eqnarray}
H &=& H_0 + H_{SO} 
\end{eqnarray}
where we define the hopping and spin-orbit coupling terms in real space as 
\begin{eqnarray}
H_0 &=& -t \sum_{\substack{\langle i j \rangle \sigma }} c_{i \alpha \sigma}^{\dagger} c_{j \beta \sigma} \\ \nonumber
H_{SO} &=& \sqrt{2} i \lambda \sum_{\substack{\langle ij\rangle \\ \alpha \beta }} \left[c_{i \alpha \sigma}^{\dagger} \frac{\bs b_{ij} \times \bs d_{ij}}{|\bs b_{ij} \times \bs d_{ij}|}\cdot \sigma_{\sigma \sigma'} c_{j \beta \sigma'} + h.c. \right],
\end{eqnarray}
respectively. Here the operator $c^{\dagger}_{i \alpha \sigma}$ creates an electron at site $i$, orbital $\alpha$ and spin $\sigma$, $\langle .. \rangle$ denotes nearest neighbor hopping  and $t$ is the nearest neighbhor hopping parameter. Further  $\lambda$ is the strength of spin-orbit coupling, $\bs b_{ij}$ is a vector that connects the center of the tetrahedron to the mid-point of the bond connecting nearest neighbor sites $i$ and $j$, and $\bs d_{ij}$ is a vector that connects nearest neighbor lattice sites $i$ and $j$. \par
\textit{Hamiltonian matrix without SOC:} Written in momentum space, the nearest neighbor hopping terms take the form
\begin{eqnarray} \nonumber
H_0 &=& \sum_{\substack{\bs k \sigma \\ \alpha \beta}} c_{\bs k \alpha \sigma}^{\dagger} h_{\alpha \beta}(\bs k) c_{\bs k \beta \sigma} 
\end{eqnarray}
where the Hamiltonian matrix $h_{\alpha \beta} (\bs k)$ is given by
\begin{eqnarray}
h_{\alpha \beta}(\bs k) &=& 
-2 t\left( \begin{array}{cccc}
 0 &  c_ {k_\text{x}+k_\text{y}} & c_ {k_\text{x}+k_\text{z}} &  c_ {k_\text{y}+k_\text{z}} \\
 c_ {k_\text{x}+k_\text{y}} & 0 & c_ {k_\text{y}-k_\text{z}} & c_ {k_\text{x}-k_\text{z}} \\
c_ {k_\text{x}+k_\text{z}} &  c_ {k_\text{y}-k_\text{z}}& 0 &  c_ {k_\text{x}-k_\text{y}} \\
  c_ {k_\text{y}+k_\text{z}} & c_ {k_\text{x}-k_\text{z}} &  c_ {k_\text{x}-k_\text{y}}& 0 \\
\end{array}
\right).
\end{eqnarray}
We have used the short hand $c_{k_i \pm k_j}$ to mean $\cos(k_i \pm k_j)$.\par
\textit{Hamiltonian matrix with SOC:} The Hamiltonian in the presence of SOC is ($ 1_{2\times 2}$ is the identity matrix)
\begin{widetext}
\beq \nonumber
H &=&  \sum_{\substack{\bs k \sigma \sigma' \\ \alpha \beta}} c_{\bs k \alpha \sigma}^{\dagger} \left[h_{\alpha \beta}(\bs k) + h_{\alpha \beta}^{SO}(\bs k) \right]c_{\bs k \beta \sigma'},  \\ \nonumber
\eeq
where the matrices $h_{\alpha \beta} (\bs k) $ and $h_{\alpha \beta}^{SO}(\bs k) $ are given as
\beq \nonumber
h_{\alpha \beta} (\bs k) &\rightarrow& 1_{2\times 2}\otimes h_{\alpha \beta} (\bs k) \\ \nonumber
h_{\alpha \beta}^{SO}(\bs k)&=& 
-2 i \lambda (\sigma_x - \sigma_y)\otimes \left( \begin{array}{cccc}
 0 &  -c_ {k_\text{x}+k_\text{y}} &0 &  0 \\
 c_ {k_\text{x}+k_\text{y}} & 0 &0 & 0\\
0& 0& 0 & 0 \\
0& 0&  0& 0 \\
\end{array}
\right) 
+
2 i \lambda (\sigma_x - \sigma_z)\otimes \left( \begin{array}{cccc}
 0 &  0 &-c_ {k_\text{x}+k_\text{z}} &  0 \\
0& 0 &0 & 0\\
c_ {k_\text{x}+k_\text{z}}& 0& 0 & 0 \\
0& 0&  0& 0 \\
\end{array}
\right) \\  \nonumber
&-&2 i \lambda (\sigma_y - \sigma_z)\otimes \left( \begin{array}{cccc}
 0 & 0&0 &  -c_ {k_\text{y}+k_\text{z}} \\
0 & 0 &0 & 0\\
0& 0& 0 & 0 \\
c_ {k_\text{y}+k_\text{z}}& 0&  0& 0 \\
\end{array}
\right)
+
2 i \lambda (\sigma_y + \sigma_z)\otimes \left( \begin{array}{cccc}
 0 &  0&0 &  0 \\
0 & 0 &-c_ {k_\text{y}-k_\text{z}}& 0\\
0& c_ {k_\text{y}-k_\text{z}}& 0 & 0 \\
0& 0&  0& 0 \\
\end{array}
\right) \\ \nonumber
&-&
2 i \lambda (\sigma_x + \sigma_z)\otimes \left( \begin{array}{cccc}
 0 &  0 &0 &  0 \\
0 & 0 &0 & -c_ {k_\text{x}-k_\text{z}}\\
0& 0& 0 & 0 \\
0& c_ {k_\text{x}-k_\text{y}}&  0& 0 \\
\end{array}
\right)
+
2 i \lambda (\sigma_x + \sigma_y)\otimes \left( \begin{array}{cccc}
 0 &  0 &0 &  0 \\
0& 0 &0 & 0\\
0& 0& 0 & -c_ {k_\text{x}-k_\text{y}} \\
0& 0& c_ {k_\text{x}-k_\text{y}}& 0 \\
\end{array}
\right), \\
&&
\eeq
where $\sigma_i$ are the Pauli matrices. 
\end{widetext}

\section{Acknowledgments}
We thank Jim Allen, Haoyu Hu, Lei Chen and Sarah Grefe for fruitful discussions. 
This research used resources of the Stanford Synchrotron Radiation Lightsource, SLAC National Accelerator Laboratory, which is supported by the U.S. Department Of Energy (DOE), Office of Science, Office of Basic Energy Sciences under Contract No. DE-AC02-76SF00515. The ARPES work at Rice University was supported by the Gordon and Betty Moore Foundation's EPiQS Initiative through grant No. GBMF9470 and the Robert A. Welch Foundation Grant No. C-2175.
The theory work at Rice has been supported by the Air Force Office of Scientific Research under Grant No. FA9550-21-1-0356 (C.S.), the U.S. Department of Energy, Office of Science, Basic Energy Sciences, under Award No. DE-SC0018197 (Q.S.), and the Robert A. Welch Foundation Grant No. C-1411 (C.S.). 
L.Z.D. and C.W.C are supported by US Air Force Office of Scientific Research Grants FA9550-15-1-0236 and FA9550-20-1-0068, the T. L. L. Temple Foundation, the John J. and Rebecca Moores Endowment, and the State of Texas through the Texas Center for Superconductivity at the University of Houston.
Y.S. was supported by the National Natural Science Foundation of China (Grants No. U2032204), and the Strategic Priority Research Program of the Chinese Academy of Sciences (Grants No. XDB33030000).
S.S. and G.C. are supported by the National Research Foundation, Singapore under its Fellowship Award (NRF-NRFF13-2021-0010) and the Nanyang Assistant Professorship grant from Nanyang Technological University. J.Y.Y. and Y.P.F. are supported by the Ministry of Education, Singapore, under its MOE AcRF Tier 3 Award MOE2018-T3-1-002.
Work at University of California, Berkeley, is funded by the U.S. Department of Energy, Office of Science, Office of Basic Energy Sciences, Materials Sciences and Engineering Division under Contract No. DE-AC02-05-CH11231 (Quantum Materials program KC2202).

\section{Author contributions}
MY oversaw the project. JH, MY and JSO carried out the ARPES measurements with the help of DL, MH, SG, YG, YZ, ZY with input and guidance from RJB. The ARPES data were analyzed by JH. Single crystals were synthesized by HL under the guidance of YS. Theoretical modeling was carried out by CS and QS. Density Functional Theory calculations with VASP were carried out by JYY and SS under the guidance of YPF and GC. Density Functional Theory calculations with WIEK2K were carried out by JDD. Transport measurements were carried out by LD and CWC. JH, CS, and MY wrote the paper with input from all co-authors.

\textbf{Competing interests:}
The authors declare that they have no competing interests.

\textbf{Data and materials availability:}
All data are available in the manuscript or the supplementary materials.

\newpage

\newpage

\begin{figure}
\includegraphics[width=0.9\textwidth]{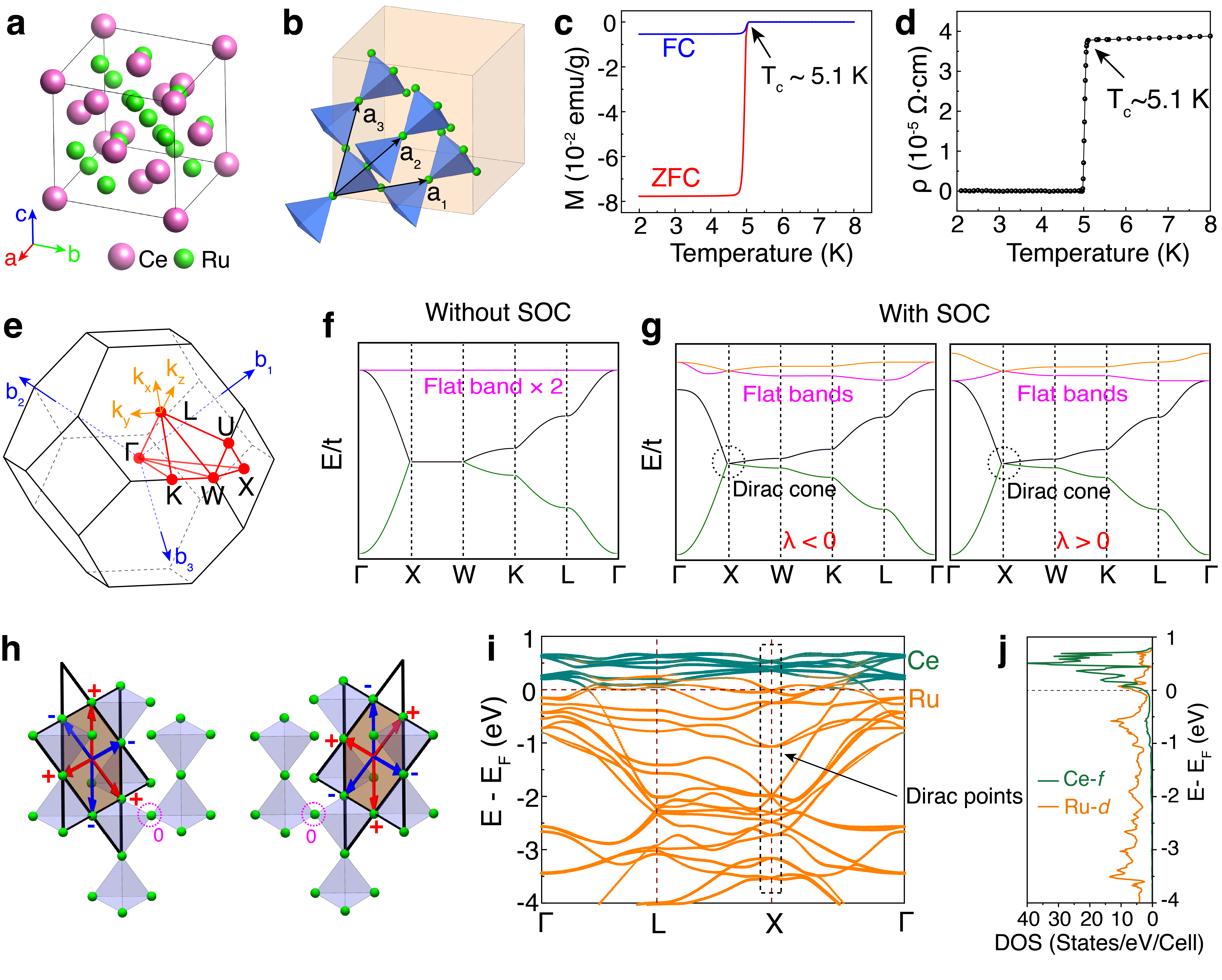}
\caption{{\bf\label{fig:Fig1}Crystal structure, destructive interference and tight-binding model.} (\textbf a) Crystal structure of \crr~shown in the conventional unit cell. (\textbf b) Ru atoms form the pyrochlore lattice, where \textbf a$_i$ are the lattice translation vectors within the faces of the cube. (\textbf c) Magnetization and (\textbf d) resistivity measurements showing superconductivity at 5.1 K. FC means field cooling and ZFC means zero field cooling. (\textbf e) Brillouin zone (BZ) with the high symmetry points labeled. We define $k_x$, $k_y$ and $k_z$ here with respect to the (111) surface for the convenience of experimental description. (\textbf f) Four-band tight-binding model of the pyrochlore lattice without considering spin-orbit coupling (SOC), where the flat band is doubly-degenerate. (\textbf g) Tight-binding model including SOC. (\textbf h) Mechanism for 3D destructive interference on the pyrochlore lattice. The sites marked by magenta circles have zero Wannier amplitude and do not produce inter-kagome-plane hopping and hence do not detract from the flatness of the bands. (\textbf i) Density functional theory (DFT) calculations with projection onto different atoms. The black dashed rectangle marks out the Dirac points at the X point of the BZ. (\textbf j) Partial density of states (DOS) of the Ce-\textit{f} and Ru-\textit{d} orbitals.}
\end{figure}

\begin{figure}
\includegraphics[width=1\textwidth]{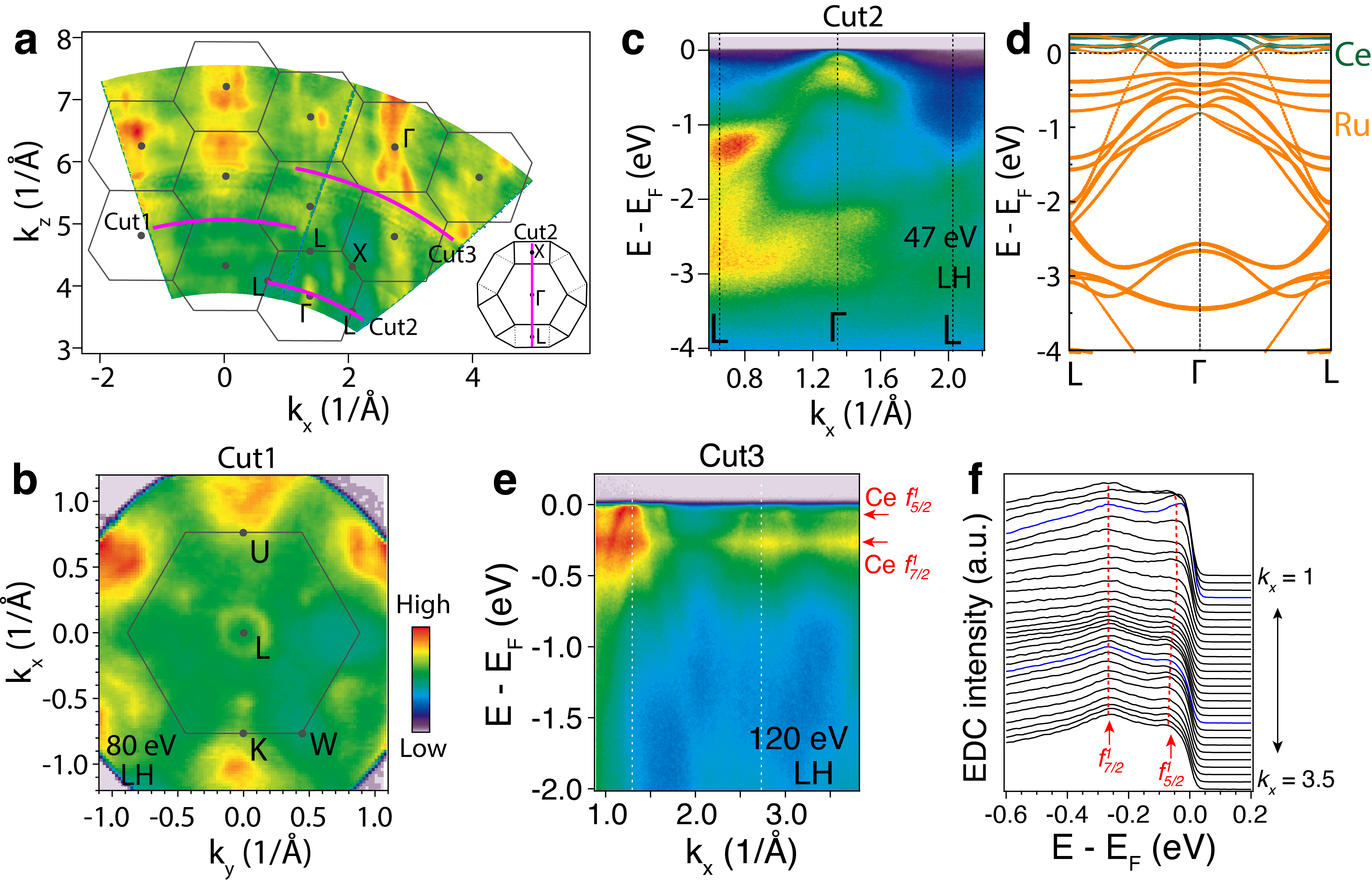}
\caption{{\bf\label{fig:Fig2}Electronic structure and Ce 4$f$ flat bands.} (\textbf a) Out-of-plane Fermi surface mapping of \crr~using photon energies from 40 eV to 200 eV. Inner potential is 22 V. (\textbf b) In-plane Fermi surface mapping with respect to the (111) cleavage surface using 80 eV photons. (\textbf c) Spectral image measured with 47 eV photons along the momentum path indicated in (a). LH indicates linear horizontal polarization. (\textbf d) DFT calculations of the band structure along the L -- $\Gamma$ -- L direction. The color and line thickness indicate the spectral weight of the different atoms. (\textbf e) Spectral image measured with 120 eV photons showing two flat bands close to \ef, which correspond to the Ce 4$f^1_{5/2}$ and 4$f^1_{7/2}$ states. The momentum path is indicated in (a). (\textbf f) Energy distribution curve (EDC) stacks of spectral image (e). The blue EDCs correspond to the momentum positions indicated by the white dashed lines in (e).
}
\end{figure}

\begin{figure}
\includegraphics[width=1\textwidth]{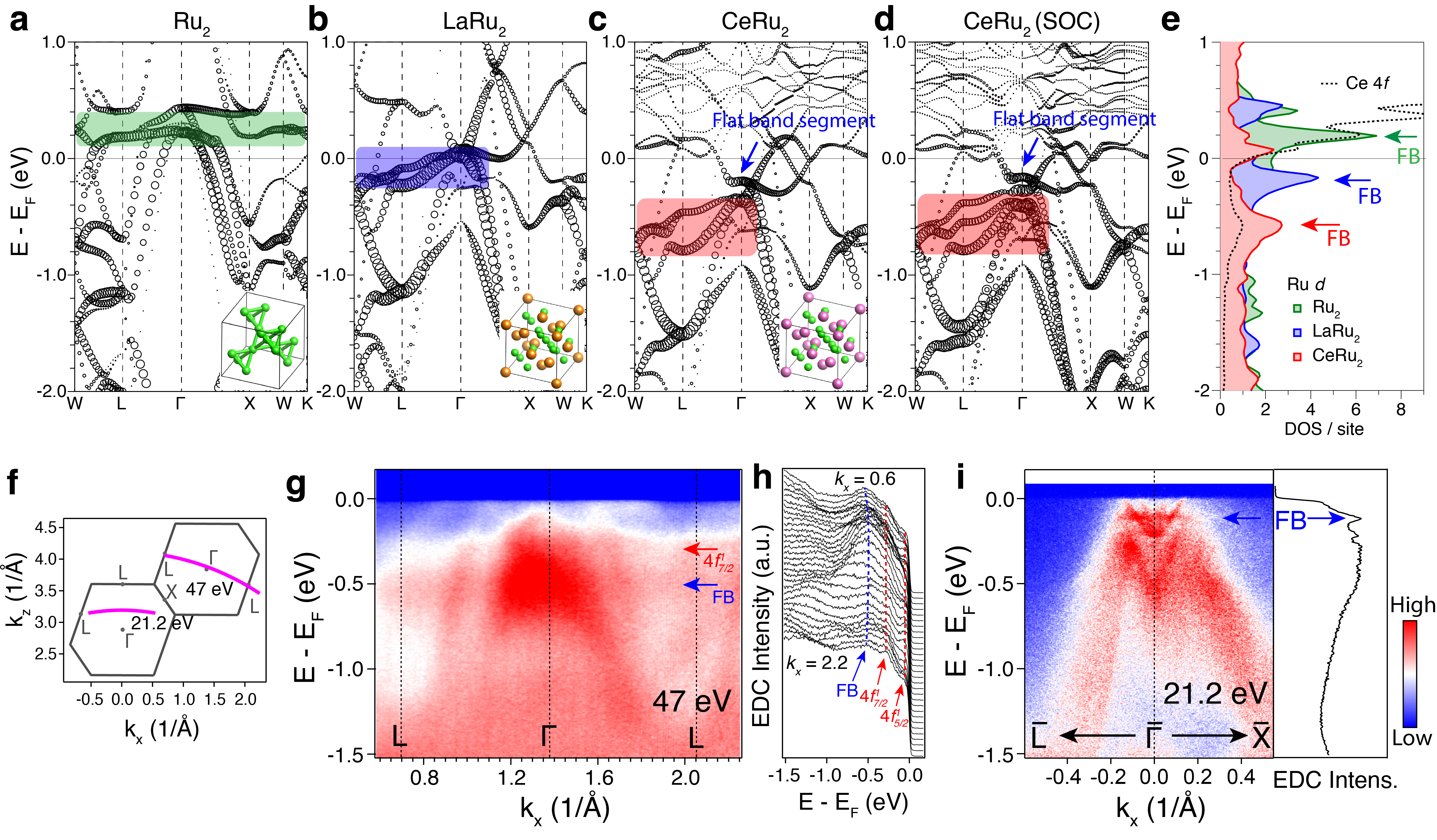}
\caption{{\bf\label{fig:Fig3}Flat bands by destructive interference.} (\textbf {a-d}) Comparison of valence band electronic structure for the (a) Ru$_2$ pyrochlore lattice only, (b) LaRu$_2$, (c) \crr~and (d) \crr~including spin-orbit coupling (SOC). The \crr~lattice constant (a = 7.53 \AA) is used for all structures. Symbol size reflects combined Ru \dz~+ \dxy~+ \dxxyy~orbital character, and the shaded rectangles indicate the bands originating from the destructive-interference-induced flat bands. (\textbf e) Comparison of k-integrated Ru-$d$ density of states illustrating the relative energy shift and dispersion broadening of the flat bands. The Ce 4$f$ DOS, primarily above \ef, also exhibits a similar hybridization peak at -0.6 eV coincident with Ru-$d$ DOS peak. (\textbf f) Side view of the BZ, with cut positions marked. (\textbf g) Spectral image along L -- $\Gamma$ -- L measured with 47 eV photons. The momentum path is indicated in (f). (\textbf h) EDC stacks of (g) showing the Ce 4$f^1_{5/2}$, 4$f^1_{7/2}$ states and flat band induced by destructive interference. (\textbf i) Spectral image and corresponding integrated EDC (-0.2 $\sim$ 0.2 \AA) measured with 21.2 eV photons. One sharp flat band is clearly revealed. The corresponding momentum path is indicated in (f).
}
\end{figure}

\begin{figure}
\includegraphics[width=0.9\textwidth]{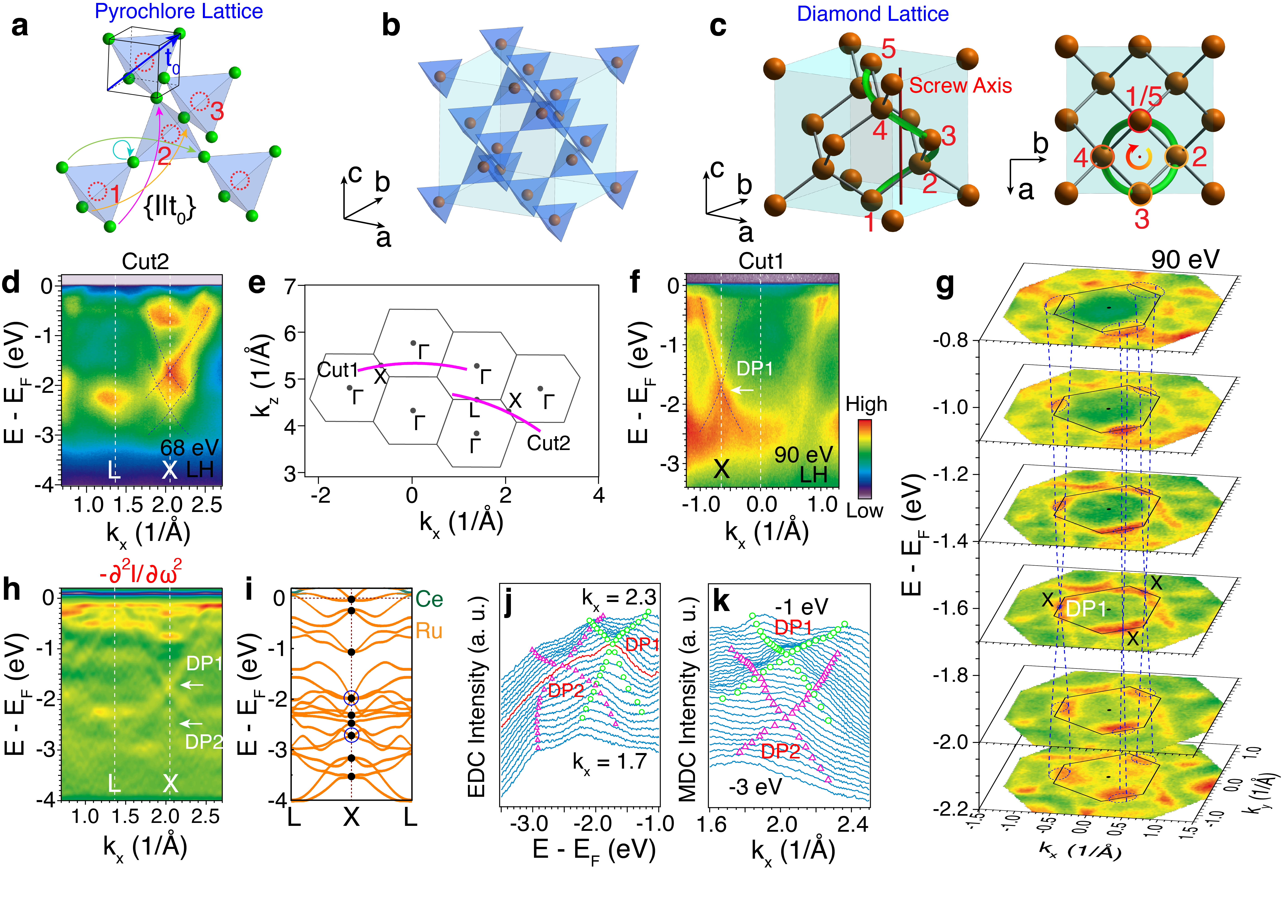}
\caption{{\bf\label{fig:Fig4}Three-dimensional Dirac cones.} (\textbf a) Pyrochlore lattice and the nonsymmorphic symmetry operation $\{I|t_0\}$, where $I$ is inversion with respect to the tetrahedron center (red circle), $t_0$ is the fractional translation along the diagonal. The arrows connecting two atoms indicate the corresponding atom translation under the operation. (\textbf b) The centers of the tetrahedra in the pyrochlore lattice form a diamond lattice. (\textbf c) The diamond lattice with marked screw axis demonstrating the nonsymmorphic symmetry protecting the Dirac crossings at X. (\textbf d) Spectral image measured with 68 eV photons corresponding to the L--X direction as shown in (e). The overlaid blue dashed lines are guidelines for the two Dirac cones. (\textbf e) Side view of the BZ, with cuts marked. (\textbf f) Spectral image measured with 90 eV photons along cut1. The overlaid blue dashed lines are guidelines for the Dirac crossing. (\textbf g) 3D view of the in-plane Dirac cone dispersion at X measured with 90 eV photons. (\textbf h) Second energy derivative of spectral image in (d). Two Dirac points (DP) are labeled by the white arrows. (\textbf i) DFT calculations of the bands along L--X--L. The black dots indicate all band crossings at X are four-fold degenerate. The corresponding two Dirac cones observed by experiments are marked out by the blue circles. (\textbf j) EDC stacks of (d) between the momentum range from k$_x$ = 1.7 \AA$^{-1}$ to k$_x$ = 2.3 \AA$^{-1}$ across the Dirac cone. (\textbf k) Momentum distribution curve (MDC) stacks of (d) between the energy range from -3 eV to -1 eV. The green circles and magenta triangles trace out the two sets of Dirac dispersions.
}
\end{figure}

\end{document}